\documentclass[prb,twocolumn,floatfix,10pt,amsmath,amssymb,showpacs,superscriptaddress]{revtex4-2}
\usepackage{amsmath,amssymb}
\usepackage{graphicx}
\usepackage{inputenc}
\usepackage{color}
\usepackage{hyperref}

\newcommand\siesta{\textsc{Siesta}}

\newcommand{\ie}{{\em i.\,e. }}

\newcommand{\bfk}{\mathbf{k}}

\newcommand{\bfv}{\mathbf{v}}
\newcommand{\Vdir}{\hat{\mathbf{e}}}
\begin{document}
	
\title{Strong current in carbon nanoconductors: Mechanical and magnetic stability}


\author{S. Leitherer}
\affiliation{Department of Chemistry, University of Copenhagen, DK-2100 Copenhagen, Denmark}
\author{N. Papior}
\affiliation{DTU Computing Center, Department of Applied Mathematics and Computer Science, Technical University of Denmark, DK-2800 Kongens Lyngby, Denmark}
\author{M. Brandbyge}
\affiliation{ Department of Physics, Technical University of Denmark, DK-2800 Kongens Lyngby, Denmark}

\begin{abstract}
Carbon nanoconductors are known to have extraordinary mechanical strength and interesting magnetic properties. Moreover, nanoconductors based on one- or two-dimensional carbon allotropes display a very high current-carrying capacity and ballistic transport.
Here, we employ a recent, simple approach based on density functional theory 
to analyze the impact of strong current on the mechanical and magnetic properties of carbon nanoconductors. We find that the influence of the current itself on the bond-strength of carbon in general is remarkably low compared to e.g. typical metals. This is demonstrated for carbon chains, carbon nanotubes, graphene and polyacetylene. We can trace this to the strong binding and electronic bandstructure.  On the other hand, we find that the current significanly change the magnetic properties. In particular, we find that currents in graphene zig-zag edge states quench the magnetism.
\end{abstract}





\maketitle
\section{Introduction}
The strong chemical bond between carbon atoms gives rise to the wide variety of mechanically strong, conducting carbon structures which are found to be remarkably stable even under strong electrical current including one-dimensional chains\cite{Romdhane2017}, nanotubes\cite{Yao2000,Javey2003}, and two-dimensional graphenes\cite{Meunier2016}. 
For graphene nanostructures the electrons may to a large degree be in the ballistic quantum transport regime which is e.g. reflected in the occurrence of quantum interference effects\cite{Caneva2018} or long mean-free paths\cite{Baringhaus2014}. The exceptional electrical conduction properties of graphene combined with its flexibility and mechanical strength makes carbon very promising for nano-scale electronic devices which also may operate at high voltages and current densities. Narrow graphene conductors show a current-carrying capacity which can reach  $10^8\,\mathrm{A}/\mathrm{cm}^2$ before breakdown \cite{Moser2007,Liu2022}, while individual carbon nanotubes appear to be able to sustain current densities even exceeding $10^9\,\mathrm{A}/\mathrm{cm}^2$.

Strong fields and currents have been used to modify nanoscale carbon materials\cite{Harris2017}.
Atomic-scale studies of graphene structures in the presence of high current and applied voltage has been performed using {\it in-situ} high-resolution transmission electron microscopy (HRTEM)\cite{Harris2017,Liu2022}. It has been seen how the structure is cleaned for residues and edges changed by the current/voltage\cite{Jia2009}, and layers fuse \cite{Barreiro2012}. The strong nonequilibrium has been applied in this way to form narrow gaps between graphene electrodes \cite{Gnr2022,ullmann2015single}. In such experiments strong current fluctuations appear, which probably is due to the formation and breaking of conducting carbon filaments down to one carbon atom in width between the electrodes\cite{Zhang2012,Sarwat2017}. Further, monatomic carbon chains -- carbyne -- between graphene has been formed and studied in HRTEM where they were observed to be stable for voltages in the 1V range and currents in the 10$\mu$A range\cite{Cretu2013}. The bonds in carbon chains possess an extreme stiffness\cite{Liu2013}. On this background it is interesting to look into what role the current plays on the strength of sp$^1$ and sp$^2$ carbon bonds.

Carbon nanostructures have also attracted significant attention due to the appearance of local magnetic moments at defects and at zig-zag edges\cite{Dimas22}, and the low spin-orbit coupling has made it interesting for spintronics applications\cite{Slota2018, Friedrich2022}. It was predicted how a graphene nanoribbon with magnetic zig-zag edges can be turned into a half-metal by applying an transversal electrical field\cite{son2006half}. For nanoelectronic devices involving the local magnetism of carbon it is interesting to address the impact of electronic current on the magnetism.

In this paper we apply a simple approach - coined {\em bulkbias}\cite{papior2022simple} -  to calculate the effects of finite electronic currents on different carbon nanoconductors.
The method is based on density functional theory (DFT), but including a nonequilibrium state occupation corresponding to a model of ballistic transport. In the method we consider periodic systems without including the effect of the electrical fields to single out the effects of current alone. Here, we use it to study how a finite electrical current modify the electronic  structure and induce changes in the mechanical strength or magnetism of some carbon nanoconductors.  We find that the influence of the current itself on the bond-strength of carbon is very low indeed. This is in stark contrast to the results found for metallic chains, where ballistic currents corresponding to voltages on the order of 1V significantly weakens the bonds. On the other hand, for magnetism we find that strong currents can quench the magnetism of graphene zig-zag edge states.

\section{Method}
In a homogeneous, ballistic nanoconductor such as a nanotube connected to metal electrodes\cite{Frank1998,Javey2003}, the resistance versus conductor length, $L$, does not follow Ohm's law,  i.e. linear increase with $L$. Instead, the resistance remains almost constant up to the mean free path for inelastic phonon exicitation\cite{Yao2000}. This is a hallmark of ballistic transport, in which case the electric field or voltage-drop is only significant around scattering centers. For systems with low defect concentration the voltage drop will concentrate around the connections to the bulk electrodes\cite{DattaBook97, Leitherer2019}. The main idea in the {\em bulkbias} method\cite{papior2022simple} is to neglect these connecting regions and the field effects altogether, and just consider the effects of a steady-state electronic current in the ``bulk'', deep inside the ballistic conductor. This situation can then be modelled as a periodic crystal lattice described by a bandstructure employing a periodic unitcell and Bloch's theorem. In order to include the current and mimic the ballistic occupation of states,  we define ``left'' and ``right'' moving states according to the projection of their band velocity along the field direction, $\Vdir$. 
To mimic the state occupation in a ballistic conductor, we fix the chemical potentials for left and right-movers relative to a quasi-Fermi level as $\mu_L=E_F+eV/2$ and $\mu_R=E_F-eV/2$ with $V$ being the applied voltage, and the field $\Vdir$ is directed from right to left. 
With this approach the effective ballistic distribution function depends on the bias and bandstructure, $\varepsilon_{n\bfk}$, with corresponding velocities $\bfv_{n\bfk}$, and is given by,
\begin{equation}
		f(n\bfk)=n_F(\varepsilon_{n\bfk}-E_F)  + \delta f_L + \delta f_R\,
\end{equation}
defining,
\begin{equation}
\begin{split}
	\delta f_L &= \Theta(+\Vdir\cdot\bfv_{n\bfk}) \left[n_F(\varepsilon_{n\bfk}-\mu_L)-n_F(\varepsilon_{n\bfk}-E_F)\right]\\
    \delta f_R &= \Theta(-\Vdir\cdot\bfv_{n\bfk}) \left[n_F(\varepsilon_{n\bfk}-\mu_R)-n_F(\varepsilon_{n\bfk}-E_F)\right] 
\end{split}
\end{equation}
Here a ``quasi'' Fermi level, $E_F$, is determined in the DFT self-consistent cycle to make the unit-cell charge-neutral, and
 $\Theta(x)$ is the Heaviside function. Note that all the described quantities will depend self-consistently on the bias, $eV$.
We note that this will result in different number of left and right moving electrons, $N_L$ and $N_R$, and that the nonequilibrium free energy\cite{Hershfield1993,Sutton2004},
\begin{equation}
\mathcal{F} = E_{\mathrm{tot}} - \mu_L N_L - \mu_R N_R\,,
\end{equation}
is used to calculate forces and stresses via the Hellmann-Feynman theorem. Contrary to an infinite, non-periodic system (e.g. device region between semi-infinite electrodes), $N_L$ and $N_R$ are well-defined, finite numbers adding up to the constant, total number of electrons in the unit cell, $N=N_L+N_R$.

For a two-dimensional system, such as graphene, we get the current density (per spin) averaged over the unit cell,
\begin{equation}
    \mathbf{j} = e\sum_n\int_{BZ}\frac{d\bfk}{(2\pi)^2} \bfv_{n\bfk}\, f(n\bfk)
\end{equation}
from which we may get the total current along $\Vdir$. For a one-dimensional (1D) system, \ie a nanotube, the corresponding expression yield the electronic current for a given bias $eV$ and in terms of the number of bands crossing a given energy, $N_b(E)$, as,
\begin{equation}
	I(V) = \frac{e}{h} \int_{-\infty}^{\infty}N_b(E) \,W(E,V)\,dE\,,
\end{equation}
where we have introduced the voltage ``window'' function,
\begin{equation}
W(E,V)=n_F(E-\mu_L)-n_F(E-\mu_R)\,.
\end{equation}
Beyond 1D systems, \ie graphene, one may use the periodicity transverse to $\Vdir$ with corresponding k-point sampling resulting in current per transverse unit cell. 

The \emph{bulkbias} method is implemented in \siesta \cite{Soler2002}.
In the following we have employed the PBE-GGA functional for exchange-correlation and the standard DZP basis-set in \siesta\cite{Garcia2020}. An optimized k-point sampling with 1000 k-points was chosen along the bias direction ($\Vdir$) according to the bias window. We employ a finite electronic temperature corresponding to room-temperature ($300$K). Our analysis was done using the
\textsc{sisl}\cite{zerothi_sisl} code. 

\begin{figure*}[tbh]
\centering
	\includegraphics[width=0.95\linewidth]{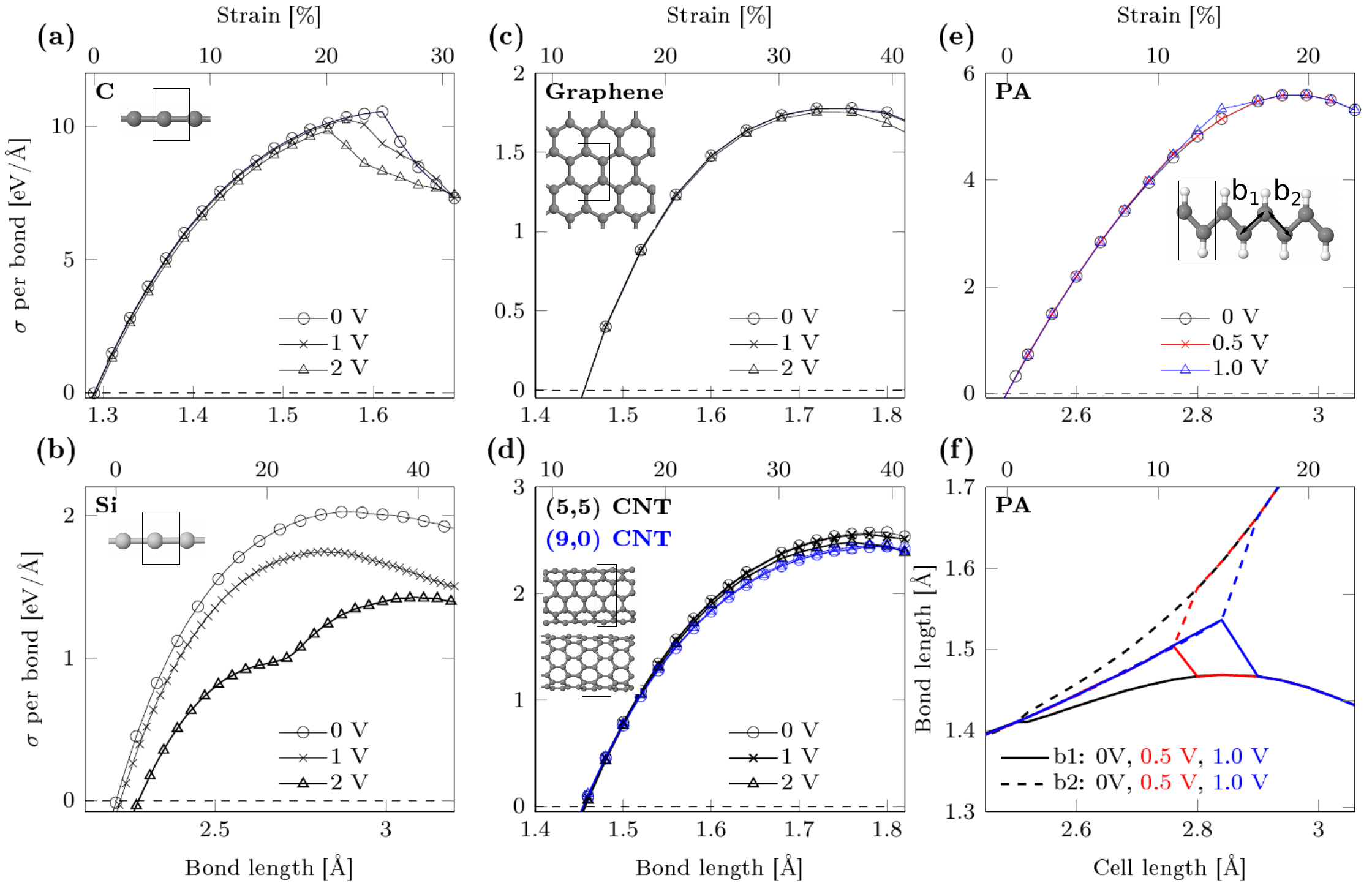}
	\caption{Stress $\sigma$ as a function of strain (bond length) applied to different carbon nanostructures. The curves are shown for different values of the bias applied in transport direction (from left to right). 
    As a measure of the size of the current, we report $I_{\mathrm{bulk}}$, which is the current at $1\,\mathrm V$ at the equilibrium cell.
    (a) Linear 1D chain of C atoms ($I_{\mathrm{bulk}}=155 \mu$A) and (b) Si atoms ($I_{\mathrm{bulk}}=232 \mu$A). The inset is depicting the structure and its 1D unit cell (not showing the vacuum regions in the transverse directions). (d) Graphene ($I_{\mathrm{bulk}}=3.24\,\mu\mathrm A/\mathrm{\AA}$),
     (c) (5,5) and (9,0) SWCNTs ($I_{\mathrm{bulk}}=155\,\mu\mathrm A$ and $I_{\mathrm{bulk}}=158\,\mu\mathrm A$), and (e) polyacetylene (PA) ($I_{\mathrm{bulk}}=76\,\mu\mathrm A$). (f)  The PA C-C bond lengths $b_1$ and $b_2$ as a function of cell length, showing the dimerization.}
	\label{fig:A}
\end{figure*}
\section{Results}

\subsection{Mechanical properties of Carbon structures}

We start by investigating the mechanic properties, \ie stress-strain curves, of several carbon-based nanostructures under influence of a \emph{bulkbias} voltage applied in transport direction, leading to a significant current flow.

Firstly, the \emph{bulkbias} method is applied to consider current in a chain of single carbon atoms, also known as carbyne. It is shown in Fig.~\ref{fig:A}a, where the inset displays the geometry and the unit cell. These have in experiments\cite{Cretu2013} been seen to be able to sustain voltages beyond 1V corresponding to currents beyond $10\mu\mathrm A$. The bulk bias method always give an upper limit to the current due to the absence of scatterers. A dimerization may be induced by the strain, however we neglect this effect here and return to it later in another context\cite{Torre2015}. We calculate the stress(force) $\sigma$ depending on the change in bond length (strain) of the chain. 
Interestingly, we find that the carbon chain is very stable up to a strain of 20\% with high maximum stresses of $\approx 10\,\mathrm{eV}/\mathrm\AA$, and, importantly, it displays only a weak bias dependence of the maximum tensile stress which is below 10\% at 2V and without significant change in bond length. This is less bond weakening compared to metal atomic chains where only Cu and Au was found to have a decrease in maximum tensile stress in this range \cite{papior2022simple}. The carbon chain is about 5 times stronger compared to the maximum stress applicable a chain of Si atoms (Fig.~\ref{fig:A}b) which is $\approx 2\,\mathrm{eV}/\mathrm\AA$ at 0 V, but, most importantly, the maximum stress and elasticity for the Si chain decrease significantly with bulk bias while the equilibrium bond length increase.

\begin{figure}[tbh]
	\centering
		\includegraphics[width=0.95\linewidth]{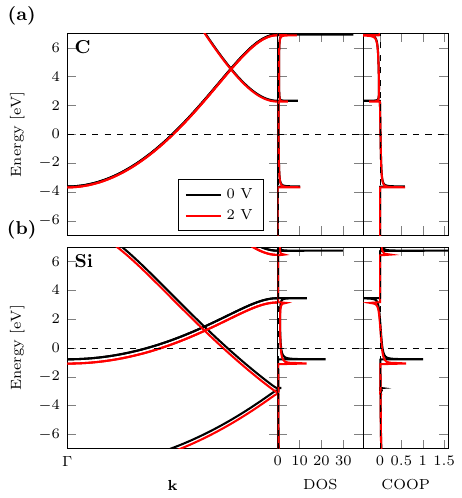}
	\caption{Band structure, density of states (DOS) and COOP for 1D C and Si chain. The quantities where extracted at a bond length of $1.5\,\mathrm \AA$ for C (a) and $2.5\,\mathrm \AA$ for Si (b). In the Si chain, the population of bonding states near the Fermi energy $E_F=0$ eV decreases, explaining its sensitivity towards the bias.}
	\label{fig:B}
\end{figure}

This behavior can be explained by the bias-dependent density of states (DOS) and crystal orbital overlap population (COOP) \cite{Hoffmann1988,papior2022simple}, where a positive/negative COOP corresponds to states with bonding/anti-bonding character. 
While the DOS of the carbon chain near the Fermi energy is low with no clear bonding/anti-bonding character and very weakly bias dependent (Fig.~\ref{fig:B}a), for the Si chain we find bonding states near the Fermi energy (Fig.~\ref{fig:B}b). As the bias increases, these Si states get depleted leading to a weakening of the bonds, \ie a reduction in stress.

The 2D infinite graphene (Fig.~\ref{fig:A}c) is treated by applying k-point sampling using 250 k-points in the transverse-to-bias direction. We applied an isotropic strain and considered the stress per bond. Again we found a very little impact of the current on the bond strength in any direction in the graphene plane, as well as a dependence on the current direction.
These results clearly are in line with the experimental fact that carbon bonds are very strong, but also very stable when a strong current is applied to the all-carbon nanostructure\cite{Moser2007,Liu2022}.

Next, we consider single-walled carbon nanotubes (SWCNTs). Here, the bulk bias is applied in the tube direction while strain is applied by stepwise increasing the bond length.
In Fig.~\ref{fig:A}d we compare a metallic zigzag (9,0) SWCNT to a (5,5) armchair SWCNT, which is semiconducting with a small bandgap of $0.09\,\mathrm{eV}$.
In order to compare more easily with the chain systems we consider the stress per bond between unitcells (see inset in Fig.~\ref{fig:A}c)
From the stress-strain curves we find that the elastic module (slope at small strains) of the (5,5) SWCNT and (9,0) SWCNT are very similar. For both SWCNTs we observe a remarkably low impact of the bulkbias, and only the stress in the (5,5) CNT decreases slightly at high bias and strain.

The same stability against strong current is also found for polyacetylene (PA). To obtain the stress-strain curves of PA, we increase the unit cell containing two atoms step-wise, and calculate the length-dependent stress in each step after a geometry relaxation (Fig.~\ref{fig:A}e). 
The stress-strain curves reveal a high stability and low bias dependency of PA.  However, the nonequilibrium influences the well-known dimerization or Peierls distortion of PA\cite{Casari2016}, which depends on the strain.
As shown in Fig.~\ref{fig:A}f, a dimerization is introduced with increasing stress, that is, we find a C-C bond length alternation, with two nonequivalent bond lengths, $b_1 \ne b_2$. Interestingly, this dimerization depends significantly on the bias. At 0V the distortion appears at a strain $>$ 2\%, while a bias of 0.5V suppresses the distortion until a strain of 10\%. 

So, in conclusion the mechanical strength is impacted very little by a strong current flowing through the C-C bonds for the various carbon systems, the carbyne chains being the most suseptible to current. But the impact of current can show up in finer details such as the dimerization of PA.

\subsection{Magnetic properties of zig-zag graphene nanoribbons}
\subsubsection{Band structure}

Graphene nanoribbons (GNRs) as carbon nanotubes are considered as candidates for nanoscale interconnects due to their high current carrying capacity. 
However, the introduction of edges in the GNRs lends the opportunity of magnetism and spintronic applications\cite{Dimas22} towards quantum information processing\cite{Wang2021}.
In this context it is interesting also to consider the influence of a finite electronic current also on the magnetic properties of  graphene nanoribbons.
Among the most prominent examples one finds GNRs with zig-zag edges (ZGNRs). These has been show to exhibit magnetism at the edge\cite{Slota2018,Blackwell2021}.
In particular, for ZGRNs in a transversal electric field, spin-polarization was predicted by DFT calculations\cite{son2006half}.
With increasing transversal electric field, semi-metallicity, i.e. metallic behavior for one spin and isolating behavior for the opposite spin, was found. In the following we will investigate, how this spin polarization is influenced when additionally to the transversal electric field, a bulk bias is applied leading to a steady-state electronic current in the ribbon.

\begin{figure}[tbh]
	\centering
\includegraphics[width=0.95\linewidth]{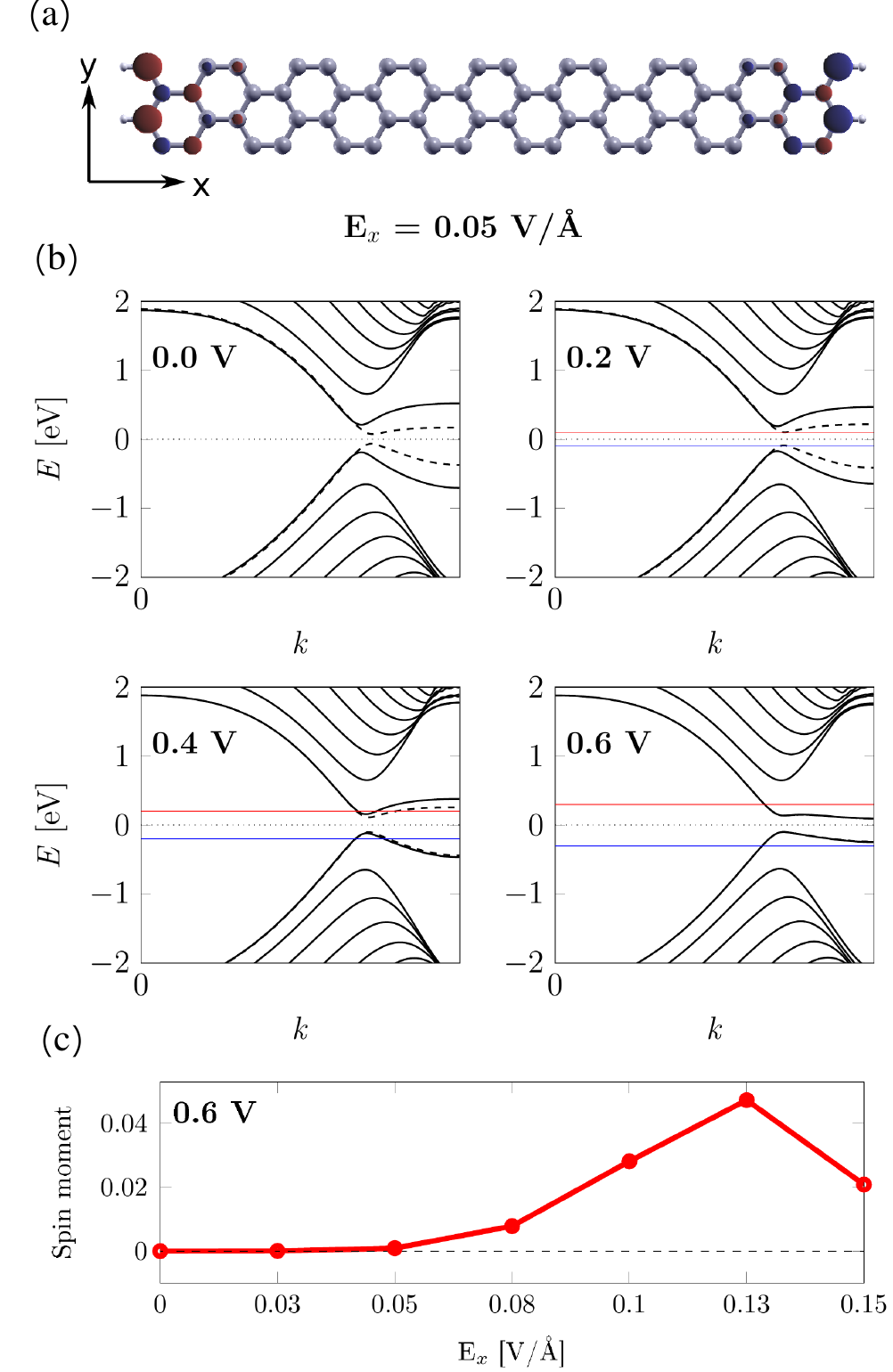}
	\caption{A zig-zag graphene nanoribbon, 16-ZGNR, in transversal electric field, $E_x$, with finite current determined by the bias applied in transport direction $y$. (a) Spin up and down density localized at the edges of the 16-ZGNR in an electric field of $E_x=0.05 \mathrm V/\mathrm \AA$ without bias. (b) Band structure under influence of bias. The voltage window is indicated by lines in red/blue color at $\mu_{L/R}$. For zero bias, there is a splitting of the spin up and down bands near the Fermi energy. The bias diminishes the spin polarization. At high bulk bias ($>0.6\,\mathrm V$), the spin splitting is quenched. (c) Spin moment at a bias of $0.6\,\mathrm V$ over transverse electric field, displaying how the spin splitting can be reintroduced by using higher fields.}
	\label{fig:C}
\end{figure}

In  Fig.~\ref{fig:C}a, we show the atomic structure of a 16-ZGNR. When brought into a fixed transversal electric field of $E_x=0.05 \textrm{\AA}^{-1}$, the bands around $E_F$ split up into a spin up and a spin down band (top left panel in Fig.~\ref{fig:C}b) in agreement with previous calculations\cite{son2006half}. The majority and minority (up and down) spin density is localized on the left and right zigzag edge, respectively. The remaining panels in Fig.~\ref{fig:C}b depict how this bandstructure changes, when a bias is applied along the ribbon ($y$), and increased up to 0.6 V. Importantly, we note that the spin polarization decreases, and a bias of $\approx 0.6\,\textrm{V}$, the spin polarization is completely quenched. This is followed by a significant change in the bandstructure.

A simple explanation for the disappearance of magnetism is that the two chemical potentials, $\mu_L$ and $\mu_R$, move away from the equilibrium $E_F$ which is located in a region of high density of states due to the edge states presenting quite flat bands. According to the Stoner criterion\cite{NoltingBook09} magnetic solutions can appear when $J\,D(E_F)>1$,  where $J$ is an exchange energy and $D(E_F)$ the density of states. In the presence of current, we will then get an effective lower DOS around the shifted, nonequilibrium chemical potentials, $(D(\mu_L) + D(\mu_R))/2\ll D(E_F)$, thus removing the magnetism.

As the magnetism grow stronger for increasing transverse field \cite{son2006half} we may also expect a higher bias and current is needed in order to quench the magnetism. This is indeed the case, as demonstrated in
Fig.~\ref{fig:C}c where the spin moment at a fixed bias of 0.6 V is show as a function of transverse electric field.

\begin{figure}[tbh]
\centering
\includegraphics[width=0.95\linewidth]{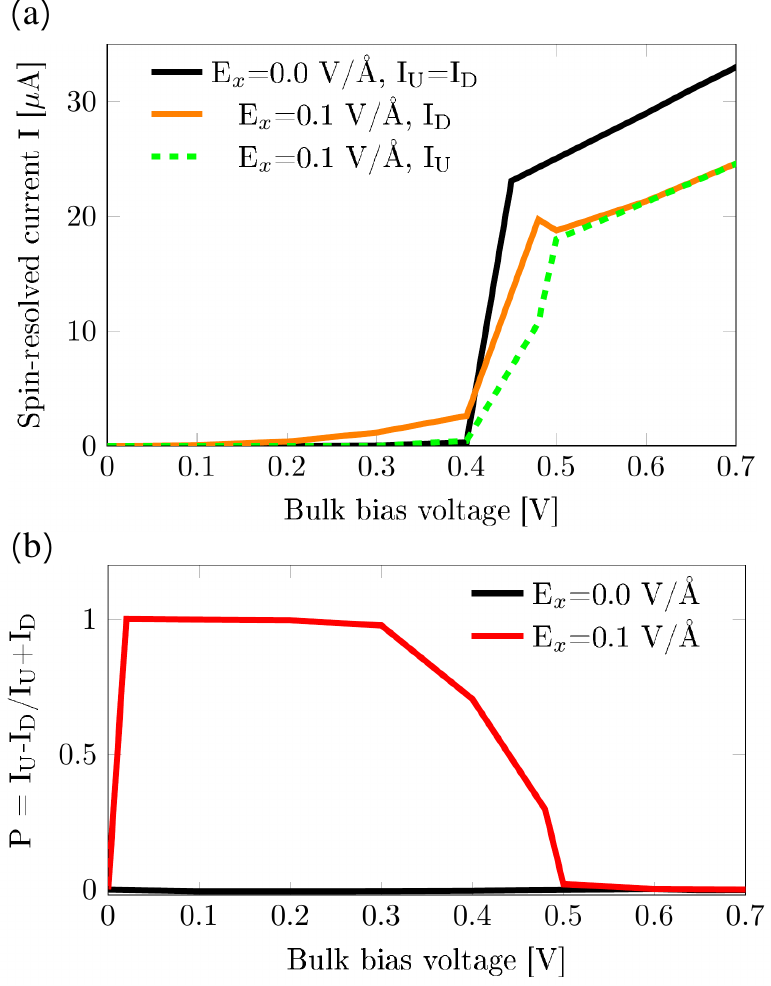}
			\caption{ 6-ZGNR in transversal electric field, $E_x$, and under a bias applied in transport direction, $y$. (a) Spin-resolved current over bulk bias voltage through 6-ZGNR without electric field (black,$I_{\mathrm{U=D}}$) and  spin current for spin up (green,$I_{\mathrm{U}}$) and down (orange, $I_{\mathrm{D}}$), with electric field $E=0.1\,\mathrm V/\mathrm\AA$ (b) Spin current polarization P for $E=0.0\,\mathrm V/\mathrm\AA$  and  $E=0.1\,\mathrm V/\mathrm\AA$. }
	\label{fig:D}
\end{figure}

\subsubsection{Spin currents}

The effect of a finite current, and the spatial current pattern, was more recently studied for narrower ZGNRs (6-ZGNRs) by Zhang and Fahrenthold\cite{zhang2021spin} using the DFT-NEGF\cite{Brandbyge2002} approach where bulk electrodes were defined at the left and right side of the ZGNRs between which a bias was applied, leading to an voltage drop across the scattering region. Here we analyze in detail the spin current flow the spin-polarized 6-ZGNRs in the presence of a transverse electric field. Contrary to the DFT-NEGF method
the {\em bulkbias} method allows for the study of the effects of nonequilibrium/current alone without the voltage drop which will depend on the details of the attachment to electrodes. 

The total spin currents through a 6-ZGNR and the spin polarization, $P$, as a function of bias are shown in Fig.~\ref{fig:D}a and b, respectively. Without electric field (black line), the spin up current $I_{\mathrm{U}}$ equals the spin down current $I_{\mathrm{D}}$, and the spin polarization is zero, as expected\cite{son2006half}. For low bias voltages, the current is very low due to the bandgap of $0.5\,\mathrm{eV}$. At $\sim0.4\,\mathrm V$ the current rises corresponding to the entry of the edge state bands into the voltage window $[\mu_R:\mu_L]$ and the Fermi smearing. 

Applying a transversal electric field $E_x=0.1\,\mathrm{V}/\mathrm{\AA}$ leads to a splitting of the spin bands. For bias voltages $\le0.5\,\mathrm V$, the spin down band is shifted closer to the Fermi level. 
Therefore, we observe a significantly larger spin down than spin up current in this range (orange and green line in Fig.~\ref{fig:D}a for $\le0.5\,\mathrm V$). This size of spin polarized currents comparable to the previously published results\cite{zhang2021spin}. However, we find for this system that at bulk bias voltages beyond  $\sim0.5\,\mathrm V$, the spin polarization of the current vanishes, $I_{\mathrm{U}}=I_{\mathrm{D}}$, as shown in Fig.~\ref{fig:D}b.

\begin{figure}[hbt]
\centering
	\includegraphics[width=0.95\linewidth]{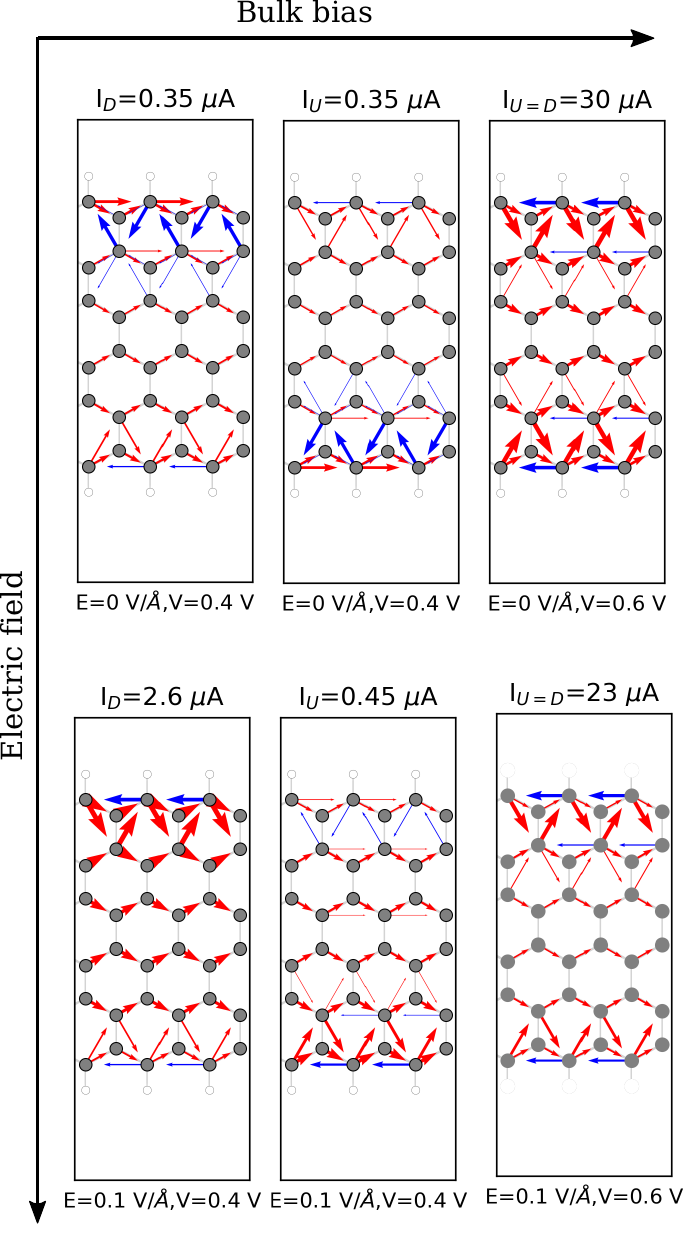}
	\caption{ Spin-resolved current (up/down spin $U/D$) patterns through a 6-ZGNR depending on the transveral electric field and on the bulk bias.  Forwards (backwards) pointing currents are depicted in red (blue).  Top row: Without electric field and for low ($0.4\,\mathrm V$) and high ($0.6\,\mathrm V$) bulk bias. Bottom row: With electric field $E=0.1\,\mathrm  V/\mathrm\AA$ and for low ($0.4\,\mathrm V$) and high ($0.6\,\mathrm V$) bulk bias. }
	\label{fig:E}
\end{figure}

The corresponding bond current patterns from our simulations are depicted in Fig.~\ref{fig:E}. The bond currents between atom $i$ and $j$ are obtained from integrating the bond transmission over the voltage window,
\begin{equation}
    I_{ij}(V) = \frac{e^2}{h}\int_{-\infty}^{\infty} t_{ij}(E,V)\, W(E,V)\,dE\,.
\end{equation}
We find that at a bias voltage of $0.4\,\mathrm V$ (Fig.~\ref{fig:E}, left and middle panels), the spin down current runs on the top and the spin down current on the bottom edge, respectively. While without transversal electric field (Fig. \ref{fig:E} top left and middle), the absolute currents are equal for both spin orientations, with a field of $E=0.1 {\mathrm V/\mathrm\AA}$ (Fig.~\ref{fig:E} bottom left and middle) they are much higher for spin down ($I_{\mathrm D}=2.6 \mu \mathrm A$, $I_{\mathrm U}=0.45 \mu \mathrm A$).
Again, these results at $V\le 0.4\,\mathrm V$ agree with what was found in \cite{zhang2021spin}. 
However, for higher bias, we find a very different behavior:
As the spin polarization vanishes, the current rises linearly with $I_{\mathrm U} = I_{\mathrm D}$.
The bond currents at $0.6\,\mathrm V$ without spin polarization (Fig.~\ref{fig:E} right panels) are now mainly localized on both edges. 
The reason for this difference is that the scattering at the junction between electrode and device regions, present in the DFT-NEGF approach \cite{zhang2021spin}, is avoided in the \emph{bulkbias} method.


\section{Conclusions}

To conclude, we have applied a recent, simple, first principles method (\emph{bulkbias} \cite{papior2022simple}) based on DFT to study the mechanical and magnetic properties of a selection of ballistic carbon-based nano-conductors under a strong current flow.  We have found that one-dimensional carbon chains (carbyne), single-walled carbon nanotubes, and poly-acetylene, as well as two-dimensional graphene all present bonds presenting a strength (maximum tensile stress) which is not weakened even in the presence of high electronic current on the order of $1-150\mu \mathrm A$ in a C-C bond. This is in accordance with the many experimental findings all pointing towards high current-carrying capacity of carbon nano-conductors\cite{Romdhane2017,Yao2000,Liu2022}.
Contrary to the mechanical stability, we have demonstrated that a ballistic current flow can considerably influence the magnetic properties predicted for ZGNRs in transversal electric fields in other studies\cite{son2006half,zhang2021spin}.
We show how the spin polarization vanishes when a sufficiently high bulk current is applied and explain this in a simple Stoner picture. The ability to switch magnetism with current may be useful for spintronics applications.
 

\section{Acknowledgements}
Funding by Villum Fonden (Grant No. 00013340). Computer infrastructure resources provided by DCC\cite{DTU_DCC_resource}.


\bibliographystyle{elsarticle-num} 

\bibliography{bulkbias.bib}

\end{document}